\documentclass[a4paper,UKenglish,cleveref, autoref, thm-restate]{lipics-v2021}

\usepackage{color,soul}

\usepackage{subfiles, tabularx, booktabs}
\usepackage{float}

\usepackage{xfp}
\usepackage{siunitx}

\newcommand{\bars}[1]{%
  \begingroup
  \setlength{\fboxsep}{0pt}%
  \strut
  {\color{gray}\rule{\dimexpr 5cm*#1\relax}{1.4ex}}%
  \endgroup
}

\newcommand{\perc}[1]{%
  \num[round-mode=places,round-precision=1]{\fpeval{100*#1/163}}%
}

\title{How Do Software Professionals Evaluate AI-Generated Code? (Registered Report)} 

\author{Samuli M\"a\"att\"a}{Software Engineering Research Group (M3S), University of Oulu, Finland
}{samuli.maatta@oulu.fi}{https://orcid.org/0009-0006-0209-4057}{FAST, the Finnish Software Engineering Doctoral Research Network, funded by the Ministry of Education and Culture, Finland}

\author{Hera Arif}{Faculty of Computer Science, Dalhousie University, Canada}{hera.arif@dal.ca}{https://orcid.org/0009-0007-6772-351X}{NSERC Discovery Grant RGPIN-2020-05001}

\author{Burak Turhan}{Software Engineering Research Group (M3S), University of Oulu, Finland}{burak.turhan@oulu.fi}{https://orcid.org/0000-0003-1511-2163}{Business Finland project MAISA (Decision number 3398/31/2024)}

\author{Paul Ralph}{Faculty of Computer Science, Dalhousie University, Canada}{paulralph@dal.ca}{https://orcid.org/0000-0002-7411-0857}{NSERC Discovery Grant RGPIN-2020-05001}

\author{Markus Kelanti}{Software Engineering Research Group (M3S), University of Oulu, Finland}{markus.kelanti@oulu.fi}{https://orcid.org/0000-0003-1886-8521}{}

\authorrunning{S. M\"a\"att\"a, H. Arif, B. Turhan, P. Ralph, and M. Kelanti} 

\Copyright{Samuli M\"a\"att\"a, Hera Arif, Burak Turhan, Paul Ralph, and Markus Kelanti} 



\category{} 

\relatedversion{} 

\acknowledgements{We thank Carolyn Seaman for advice on developing our positionality and reflexivity statement, and Tuure Tuunanen and Juuli Lumivalo for advice on laddering interviews.}

\nolinenumbers 

\begin{document}

\maketitle

\begin{abstract} 
  Recent advances in generative AI tools have significantly changed how software professionals write, evaluate, and interact with code. Generative AI tools such as GitHub Copilot, ChatGPT, and Claude are increasingly being integrated into everyday workflows. Despite the growing adoption of and reliance on these tools, it remains unclear as to \textit{how} software professionals evaluate the code they generate. To explore this topic, we will conduct a \textit{constructivist grounded theory} study that incorporates a survey, semi-structured interviews, and laddering interviews. With the initial survey data collection complete, we aim to interview 20--50 software professionals iteratively until theoretical saturation is achieved. This research aims to build a theory of how software professionals evaluate AI-generated code, grounded in their accounts of evaluative practices, perceptions, and preferences.
  \end{abstract}

\section{Introduction}

Professional software developers are increasingly adopting and relying on generative AI-based tools and agents in their work. The early literature on the topic suggested a shift from writing to reading, evaluating, and repairing generated code (e.g., \cite{bird,  between, like}). These tasks are not always easy:
\begin{itemize}
    \item not being the author of the code may make it more difficult to understand \cite{expectation} 
    \item it can contain bugs that are hidden or different from those authored by humans \cite{like, validating}
    \item it can be more difficult and time-consuming to debug and repair \cite{copilot, large-scale, like, expectation}
    \item the expectations of users do not always align with the tool's capabilities \cite{copilot, small-step}
\end{itemize}
There are also broader issues such as tendencies to outsource critical thinking and comprehension effort, and the risks of automation bias, over-reliance, skill decay, and impaired learning \cite{ironies, copilot, critical, large-scale, autopilot, skill-formation, gaironies, expectation}. Reading code requires time and mental effort. Direct forms of evaluation require investing time and effort in code comprehension, which developers tend to avoid whenever task completion is a priority \cite{comprehension}. Generative AI provides ways to use more indirect forms of evaluation (e.g., asking AI to critique or test code) and to take shortcuts in comprehension and evaluation by offloading them to AI. In particular, the issue of over-reliance has been frequently reported in previous work \cite{readable, copilot, live, large-scale, understanding, its-weird, assistant, expectation}.




The use of AI-generated code is becoming increasingly prevalent in development settings shaped by productivity pressure from sources such as deadlines, competition, and rising expectations, alongside other practical constraints, including incomplete understanding, contextual ambiguity, and organisational policies. Software professionals must increasingly decide when to evaluate, rely on, reject, regenerate, or repair AI-generated code under these conditions while preserving the productivity gains the tools promise. However, such gains may be negated when careful evaluation requires dedicating time to reading, understanding, testing, regenerating, and repairing the code. As generative AI-based tools and agents become more capable in software development tasks, it may also become more difficult for software professionals to keep up with evaluating their outputs, decisions, and actions. Together, these conditions may create an incentive to rely on generated code with less attentive evaluation, increasing the risk of unwarranted trust and over-reliance.

These challenges remain important as long as human effort is still required to critically judge AI-generated code, and they extend beyond functional correctness to the code's quality, alignment with intent and requirements, implementation and design decisions, and how professionals assume responsibility for all of these.

In this study, the research objective is to develop our understanding of these challenges from the perspective of software professionals. To this end, we start with an in-depth investigation of the perceptions, preferences, and practical conditions that shape software professionals' evaluative practices in AI-assisted programming (assessment of the value of generated code, and potentially the decisions and actions of AI agents that lead to it), as well as the characteristics of these practices, 
guided by the following initial research question: 
\textbf{\textit{RQ:~How do software professionals approach the evaluation of AI-generated~code?}}




This initial RQ is intentionally broad to keep the direction of the investigation open and let it evolve as the study progresses and questions of greater significance emerge \cite[p. 25]{charmaz}. To answer the RQ, the initial plan is to interview approximately 20--50 software professionals who have substantial first-hand experience in programming with generative AI. The target range is also intentionally broad because we plan to conduct two types of interviews (see Section \ref{met}), and the final number of participants will depend on when theoretical saturation is achieved. The initial interview design will be informed by an analysis of responses to a recent survey on software professionals' use of generative AI in Finland.

Our data analysis and collection are guided by the constructivist grounded theory (GT)~\cite{charmaz} as it aligns with our epistemological stance on qualitative research, which holds that knowledge is constructed rather than discovered (see Section \ref{section:stance}). It also makes us attend more closely to how the research participants themselves construct the present topic.

While we acknowledge the importance of observing and measuring characteristics of evaluative practices, such as assessments of code quality, we place more importance on understanding how software professionals construct and reason about the evaluation of AI-generated code as the technology and its use are rapidly evolving. For example, what counts as sufficiently careful evaluation under practical constraints such as deadlines or incomplete understanding, or who is seen as responsible for problems originating from AI-generated outputs (and whether improvement is expected from the tool, the developer, or the surrounding review practices) are questions that cannot be answered without attending to the subjective experiences of those involved. Hence, using in-depth interviews as the primary method from a constructivist perspective seems suitable for answering the research question.

Before interviews, we conducted a non-exhaustive review of the existing literature to identify research gaps. The review was enough to reveal a research gap, or rather, a research need to do a more in-depth investigation. The closest prior work, and an inspiration for our topic, was \textit{Grounded Copilot: How Programmers Interact with Code-Generating Models}~\cite{copilot} where Barke et al. identified programmers' validation strategies, including ``\textit{code examination, code execution (or testing), relying on IDE-integrated static analysis (e.g. a type checker), and looking up documentation}'', as well as validation by ``\textit{pattern matching}''. These strategies form part of their theory regarding programmers' interaction modes with GitHub Copilot. We also identified a few other grounded theory studies in adjacent topic areas (e.g., \cite{li2024, tightrope}). For example, Li et al.~\cite{li2024} conducted a socio-technical grounded theory (STGT) study that uncovered individual and organisational motives, relationships, and challenges impacting software professionals' adoption of AI tools.


Existing studies offer useful insights, but there remains much more richness in the phenomenon to be uncovered, as well as a need in the literature for a cohesive, evidence-based theory of evaluation practices themselves.

\section{Research Methodology} \label{met}

This section outlines the approach to data collection and analysis processes. Our approach is guided by the grounded theory method, specifically the constructivist variant \cite{charmaz}. We identified interviews as an appropriate method to start addressing the research question. The future direction of this study, including the choice of methods and research sites, is shaped by the emerging grounded theory analysis \cite{charmaz} (see Section \ref{section:analysis}), although prioritising interviews is preferred.









To begin this investigation, a survey was conducted targeting software professionals working in Finland. The survey addressed topics such as the extent of generative AI adoption and perceived effects. It also included questions about the professionals' evaluation-related practices and trust. The full survey results will be reported separately; here, we focus on the parts of the survey that form the base for exploration in the present study (Section \ref{section:survey}).

To gain a deeper understanding of how software professionals approach the evaluation of AI-generated code, we continue by investigating how they construct and reason about their evaluative practices. For this purpose, laddering interviews provide a suitable starting point, as they allow us to investigate why professionals prefer or avoid certain practices, how they perceive their consequences, and what value structures or goals underlie these perceptions and preferences (Section \ref{section:laddering}). We plan to complement the laddering interviews with semi-structured interviews to support more open-ended exploration and to pursue emerging directions that laddering interviews cannot effectively capture as the analysis develops.


\subsection{Initial Survey} \label{section:survey}

We recently conducted a survey on the use of generative AI by software professionals in Finland, with 163 responses collected from December 2025 to February 2026. The questionnaire consisted of five parts: (i) generative AI tools in software engineering activities, (ii) generative AI in programming, validation practices for generated code, and trust, (iii) effects of generative AI on work, (iv) background questions, and (v) closing questions.

Part (ii) serves as the basis for our investigation. It included the following three optional open-ended questions:

\begin{enumerate}
    \item \textit{In general, do you validate AI-generated code differently from human-written code? How? (For example, are there some common weaknesses you often look for?)} (51 responses)
    \item \textit{What practices help you stay productive with generative AI while managing the risks related to code quality?} (79 responses)
    \item \textit{Since you started using generative AI tools, has your trust in them changed in any way? If so, why?} (103 responses)
\end{enumerate}

These open-ended responses will be analysed following the constructivist GT approach (see Section \ref{section:analysis}). Part (ii) included ten additional closed-ended questions concerning, for example, the scope of generation (size and comprehension effort), validation practices (such as validation methods and timing), and trust in code generation tools. Although these closed-ended responses are not the main focus of the analysis, their quantitative analysis may nonetheless provide useful insights for the GT analysis.

Other parts of the questionnaire may similarly be used when relevant. For example, those who reported using generative AI in code reviews in part (i) were asked to briefly describe how. In part (v), respondents were asked about their broader thoughts and predictions about the future role of AI in software development, and some responses touched on evaluation. The other parts of the questionnaire also provide contextual information to aid the analysis, including demographic information, such as the years of experience in software development and roles or job titles (see Table \ref{demographics}), as well as information on the frequency of using generative AI and AI-generated code in software development, and the programming languages used with generative AI.

The questionnaire was available in both Finnish and English versions. It was designed and piloted iteratively until the pilot participants reported no difficulties in understanding or answering the questions or issues in translation quality. We recruited participants through convenience and snowball sampling using personal and professional social networks, LinkedIn advertisements, and various email lists and chat groups as recruiting channels, and by encouraging our contacts to forward the survey invitation to potential participants among their acquaintances.


\begin{table}[h]
\begin{tabularx}{\linewidth}{r r r X}
\toprule
\textbf{Characteristic} & \textbf{$\mathrm{n}$} & \% \\
\midrule
Years of experience \\
\midrule
< 10   & 79 & \perc{79} 
& \bars{79/163} \\
10--19 & 40 & \perc{40} 
& \bars{40/163} \\
20--29 & 23 & \perc{23} 
& \bars{23/163} \\
30+    & 21 
& \perc{21} & \bars{21/163} \\
\midrule
Role (job title) \\
\midrule
Full-Stack Developer & 37 & \perc{37} 
& \bars{37/163} \\
Software Architect & 15 & \perc{15} 
& \bars{15/163}  \\
Back-End Developer & 12 & \perc{12} 
& \bars{12/163} \\
DevOps Developer & 11 & \perc{11} 
& \bars{11/163}  \\
Data Analyst / Data Engineer / Data Scientist & 10 & \perc{10} 
& \bars{10/163} \\
Development Manager & 7 & \perc{7} 
& \bars{7/163} \\
CEO / CIO / CTO & 7 & \perc{7} 
& \bars{7/163} \\
Front-End Developer & 7 & \perc{7} 
& \bars{7/163} \\
QA / Test Engineer & 6 & \perc{6} 
& \bars{6/163}  \\
Product Owner / Product Manager & 5 & \perc{5} 
& \bars{5/163}  \\
Project Manager & 4 & \perc{4} 
& \bars{4/163}  \\
UI / UX Designer & 3 & \perc{3} 
& \bars{3/163}  \\
Business Analyst & 2 & \perc{2} 
& \bars{2/163}  \\
Enterprise Architect & 2 & \perc{2} 
& \bars{2/163} \\
Other & 35 & \perc{35} 
& \bars{35/163}  \\
\bottomrule
\end{tabularx}
\caption{Software development experience and roles in the initial survey sample}
\label{demographics}
\end{table}


A note on terminology: In the initial survey, the term \textit{validation} was used instead of \textit{evaluation}. The term was adopted from the Grounded Copilot study \cite{copilot}, where the authors used it ``\textit{broadly, to encompass any behavior meant to increase user’s confidence that the generated code matches their intent.}'' The same definition was used in our questionnaire. In a related large-scale survey \cite{large-scale}, the questionnaire items based on the findings of Grounded Copilot study used the term \textit{evaluation} instead.\footnote{The related literature does not use terminology consistently; for example, \cite{between} uses the verbs \textit{verify} and \textit{double-check}, \cite{bird} uses \textit{assess}, and \cite{readable} uses \textit{inspect}.} However, \textit{validation} was found in survey piloting to be more easily understood, as it conveys more strongly the idea of confirming that the generated code is valid.

In this GT study, our interest is initially broader and directed at all practices that software professionals use to evaluate or assess the value of AI-generated source code-related outputs or actions (e.g., their functional correctness, quality, alignment with intent or requirements, and trustworthiness) without limiting our interest to practices confirming their validity/conformance to user needs (via validation), or veracity/conformance to specifications (via verification) in any strict sense. In this study, we will use the term \textit{evaluation}, although as the GT emerges, some different term may later prove more appropriate.

\subsection{Interview Design and Piloting} \label{section:laddering}
This study will use traditional, semi-structured interviews in conjunction with laddering, which is an in-depth interview technique \cite{laddering} used as an initial data collection method. The method has been used in consumer research to elicit consumers' ``ways of thinking'' regarding a product or service category \cite{laddering} (see \cite{micro-level} for an example). It involves the use of repeated \textit{why} questions as probes to move from concrete attributes (typically of a product), or rather, in this context, means (e.g., using/not using AI to review AI-generated code), to the higher-level consequences of those means. For example, asking \textit{``Why is it important to you not to use AI for reviewing AI-generated code?''} can give insights into the consequences perceived by the interviewee (e.g., \textit{reduced code awareness}). Ultimately, this technique uncovers the personal relevance of these consequences in the form of values or goals which in this example could be \textit{professional integrity} (a value is reached when the interviewer notices that the why-questions have reached an ``end'', which is sometimes better described as a ``goal'' rather than a value).



In this study, the identified attribute-consequence-value chains (i.e., means-end hierarchies \cite{laddering}) serve as the basis for theorising using the GT method. With an initial focus on how practices (means) relate to values (ends), we may uncover interesting patterns, e.g., \textit{where do programmers prefer to invest evaluation effort?} or \textit{how do programmers approach situations in which careful evaluation conflicts with demands for increased productivity?} The identified values and goals can be assumed to be somewhat stable underlying drivers of practices. Practices can also be shaped by company policies, goals, and values, which can likewise be identified using laddering. Although concrete instances of evaluation practices vary across contexts, there may be value in knowing why programmers tend to prefer or avoid specific approaches. This can, for example, be useful for predicting how these practices change as generative AI-based tools improve.

The interviews will be conducted in Finnish and English. They are expected to last an average of 40 minutes. Each interview starts with a list of stimuli, which includes a few scenarios of what generative AI can do, from straightforward code generation to more autonomous actions. The scenarios are used to examine how participants construct and reason about evaluation under varying practical conditions, initially focusing on scenarios where it is challenging for humans to keep up. The findings of our initial survey can be used as a basis for the scenarios (see Section \ref{section:survey}). If the survey data are insufficient for developing sufficiently relevant or varied scenarios, the list may also be compiled by synthesising responses to a pre-interview survey sent when contacting potential participants.

At the beginning of each interview, the stimuli are ranked by the interviewees according to their relevance. For the highest-ranking stimulus, the interviewee is asked to relate the stimulus to concrete experiences from their own work, and to come up with a few scenarios they consider representative. For each scenario, we can ask questions such as: ``\textit{What would you evaluate in this?}'', ``\textit{What kinds of evaluation strategies would you use?}'', or ``\textit{How do you make sure you can trust the output?}''
The actual questions must be refined by piloting the interview guide. The answers are the \textit{attributes}/\textit{means}, which are then followed by a series of why-questions. Attribute-consequence-value chains are created for all attributes listed by the interviewee. These chains will be integrated into the constructivist GT analysis by coding the attributes, consequences, values, and links between them, comparing chains within and across participants, and using these comparisons alongside semi-structured interview transcripts and the initial survey responses to develop GT categories.


The interview guide is expected to be revised as the analysis continues (cf. \cite{gt-lessons}). Although the stimuli list does not change in a typical laddering study, it may change in this study if new directions emerge during the grounded theory analysis; developing the theory is the priority, not comparability and compatibility of individual attribute-consequence-value chains.

The laddering is followed by a ranking of the chains and background questions. The chain ranking is based on the relevance of the chains for the interviewees and provides useful information for analysis. This also gives the interviewees an opportunity to provide feedback on their accuracy. Background information such as software development experience and familiarity with AI tools will provide contextual information for the analysis. Each interview concludes with a question about the interviewee's interest in being contacted for more questions about the topic or participating in future research.

During the laddering interview, the interviewer writes down the interviewee's statements in a condensed and structured format to facilitate the interview. An audio recording device is used to verify the accuracy of the written information after the interview. In addition, the transcripts can be used in data analysis if they are found to contain insights that cannot be captured by the chains. All files will be uploaded either to NVivo\footnote{https://lumivero.com/products/nvivo/} or Quirkos\footnote{https://www.quirkos.com/} for analysis.

\subsection{Sampling}

Initially, we plan to recruit participants from the 66 respondents who expressed openness to being contacted for additional discussion on the survey topic by providing their email addresses. 44 potential interviewees remain after excluding the software professionals working in academia (19 out of the 163) and those not using (or only rarely using) AI-generated code in their software development work. Since all survey participants were from Finland, we aim to maximise the demographic diversity of the initial interview sample during the first round of data collection. Following the GT approach, the subsequent rounds are more concerned with what is useful for the emerging theory than with generalisability. That is, the main concern of theoretical sampling is to saturate emerging theoretical categories and their properties, not statistical generalisability. Finding potential interviewees from the survey sample is the priority, but the sampling frame is not restricted to it.




\subsection{Constructivist Grounded Theory} \label{section:analysis}


This study follows the constructivist grounded theory method \cite{charmaz}. The core features of the GT method include simultaneous data collection and analysis, theoretical sampling, theoretical sensitivity, coding, memo-writing, constant comparison, and memo sorting, with the intention of reaching theoretical saturation and cohesive theory \cite{gt}. 
Data collection will be performed until theoretical saturation is achieved, i.e. until no new themes or insights emerge from iterative data collection as the emerging theory is sufficiently supported by the collected data \cite{gt}.
Charmaz's constructivist reinterpretation of the method is one of the available variants, most notably Glaser's GT \cite{glaser1, glaser2} and Strauss and Corbin's GT \cite{strauss}.

Survey data, laddering chains, and interview transcripts provide the data that is analysed using the constructivist GT method. The first step is initial coding, followed by focused coding and theoretical coding. Initial coding refers to analytically examining and labelling data word-by-word, line-by-line, or incident-by-incident, while remaining open to the theoretical possibilities suggested by the data (not injecting the researcher's assumptions, biases, or motivations) \cite{charmaz, gt}. In focused coding, the most significant or frequent initial codes are used to categorise the data. 

Constant comparison and memo-writing are also integral parts of the GT data analysis. Constant comparative methods are used to make comparisons at each level of the analysis process, to find similarities and differences between data (e.g., statement-to-statement comparisons within the same or different interviews), codes, memos, and categories \cite{charmaz, gt}.
Memo-writing refers to writing analytic notes (or diagrams, sketches, etc.) about the codes, data, and theoretical categories as they emerge throughout the research process \cite{charmaz, gt}. They serve diverse purposes, including capturing important ideas and hunches as they appear, documenting the analysis, discovering gaps in data collection, demonstrating connections between categories, and providing material for writing the research article.

\subsection{Research Philosophy} \label{section:stance}

The main reason for choosing the constructivist version is that we tend to agree with Charmaz's stance, including the view that grounded theories are constructions of reality rather than exact pictures of it. The researcher is not a passive, value-free, and neutral observer that ``discovers'' facts from data---``their values shape the very facts that they can identify'' \cite{charmaz}. Therefore, the researcher is challenged to be more sensitive about what they inevitably bring to the research setting and how this influences their findings.

However, for this study, we do not fully commit to social constructivism, at least in any strong sense. Charmaz also encourages flexible engagement with the method rather than enforcing rigid prescriptions, e.g., with respect to epistemological stances \cite[p. 316]{charmaz}. We see the value of the constructivist lens in acknowledging how meanings and analyses are constructed, and we let it inform the epistemological stance for this study. The ontological stance for this study is more aligned with that of critical realism. To be more transparent about our position, our research interest extends beyond these constructions to the underlying mechanisms. Although it is difficult to say at this point where the theory will lead, our preliminary aim is to study a reality that is (at least partially) independent of these constructions, yet suggested by them---even though the resulting view of this reality will inevitably be a construction itself, imperfect and probabilistic.

\subsection{Positionality and Reflexivity}

The first author is a PhD researcher whose understanding of software engineering and generative AI has been developed through formal education, research, and personal projects. His use of generative AI tools, while frequent, remains cautious and selective. As he has not worked as a software professional in industry, he begins the investigation of the phenomenon from an informed and curious but somewhat distanced position. Although he acknowledges that applications of generative AI can have clear benefits in software engineering, he treats all narratives and claims about generative AI as suspicious. He is particularly concerned with how AI is used uncritically to displace, devalue, or obscure human skill, judgement, and autonomy, and with the threat of professionals and organisations becoming dependent on AI providers. This positionality makes him pay special attention to the distinctions between actual and ostensible benefits, short-term and long-term consequences, and the narratives that users tell each other and themselves.

The second author is a PhD researcher who has performed extensive research on code quality measurement, their standards, and how these measurements are validated. She also has research on challenges and issues faced while applying code quality metrics to assess code quality which could influence her choice of questions in the interview guide, as she could be unintentionally comparing the practices and challenges faced in human code evaluation to those faced during agentic code evaluation.
Her experience in the software development industry is limited and dates back approximately a decade. Her exposure to AI code generation has primarily been through hobby projects and the generation of small, preliminary code snippets within larger projects.
She believes that, despite the advantages of AI assisted development especially in terms of time saved, AI cannot replace human written code. While she considers the use of AI for generating small code snippets acceptable, she would not trust or feel comfortable relying on AI for project level code generation. Consequently, she approaches AI generated code with caution and subjects it to careful scrutiny.

The first two authors are responsible for the collection and analysis of the data, while the remaining authors provide advice and feedback throughout the process. Individually, we are committed to seeking interpretations and perspectives that challenge our own thinking and preconceptions. As a team, we will manage reflexivity by having frequent meetings to review each other's transcripts, codes, and memos, to discuss the coding process and emerging theory, and to scrutinise them for any biases related to our preconceived opinions and experiences on the topic. In addition, as recommended by Charmaz \cite[p. 166]{charmaz}, we plan to use methodological journaling to document ``methodological dilemmas, directions, and decisions'', as a way to engage in reflexivity.

\bibliography{REFS}

@inproceedings{gt,
author = {Stol, Klaas-Jan and Ralph, Paul and Fitzgerald, Brian},
title = {Grounded theory in software engineering research: a critical review and guidelines},
year = {2016},
isbn = {9781450339001},
publisher = {Association for Computing Machinery},
address = {New York, NY, USA},
url = {https://doi.org/10.1145/2884781.2884833},
doi = {10.1145/2884781.2884833},
booktitle = {Proceedings of the 38th International Conference on Software Engineering},
pages = {120–131},
numpages = {12},
location = {Austin, Texas},
series = {ICSE '16}
}

@INPROCEEDINGS{gt-lessons,
  author={Sedano, Todd and Ralph, Paul and Péraire, Cécile},
  booktitle={2017 IEEE/ACM 5th International Workshop on Conducting Empirical Studies in Industry (CESI)}, 
  title={Lessons Learned from an Extended Participant Observation Grounded Theory Study}, 
  year={2017},
  volume={},
  number={},
  pages={9-15},
  doi={10.1109/CESI.2017.2}}

@article{laddering,
	title = {LADDERING THEORY, METHOD, ANALYSIS, AND INTERPRETATION.},
	volume = {28},
	issn = {0021-8499},
	doi = {10.1080/00218499.1988.12467766},
	language = {eng},
	number = {1},
	journal = {Journal of Advertising Research},
	author = {Reynolds, Thomas J. and Gutman, Jonathan},
	month = feb,
	year = {1988},
	note = {Publisher: Ascential Events (Europe) Limited},
	pages = {11--31},
}

@book{charmaz,
  title={Constructing Grounded Theory},
  author={Charmaz, Kathy},
  edition={3rd},
  year={2024},
  publisher={Sage}
}

@article{copilot,
author = {Barke, Shraddha and James, Michael B. and Polikarpova, Nadia},
title = {Grounded {C}opilot: How Programmers Interact with Code-Generating Models},
year = {2023},
issue_date = {April 2023},
publisher = {Association for Computing Machinery},
address = {New York, NY, USA},
volume = {7},
number = {OOPSLA1},
url = {https://doi.org/10.1145/3586030},
doi = {10.1145/3586030},
journal = {Proc. ACM Program. Lang.},
month = apr,
articleno = {78},
numpages = {27}
}

@article{micro-level,
author = {Tuure Tuunanen and Juuli Lumivalo and Tero Vartiainen and Yixin Zhang and Michael D. Myers},
title ={Micro-Level Mechanisms to Support Value Co-Creation for Design of Digital Services},

journal = {Journal of Service Research},
volume = {27},
number = {3},
pages = {381-396},
year = {2024},
doi = {10.1177/10946705231173116},
URL = { 
        https://doi.org/10.1177/10946705231173116
}
}

@inproceedings{small-step,
author = {Ferdowsifard, Kasra and Ordookhanians, Allen and Peleg, Hila and Lerner, Sorin and Polikarpova, Nadia},
title = {Small-Step Live Programming by Example},
year = {2020},
isbn = {9781450375146},
publisher = {Association for Computing Machinery},
address = {New York, NY, USA},
url = {https://doi.org/10.1145/3379337.3415869},
doi = {10.1145/3379337.3415869},
booktitle = {Proceedings of the 33rd Annual ACM Symposium on User Interface Software and Technology},
pages = {614–626},
numpages = {13},
location = {Virtual Event, USA},
series = {UIST '20}
}

@inproceedings{expectation,
author = {Vaithilingam, Priyan and Zhang, Tianyi and Glassman, Elena L.},
title = {Expectation vs. Experience: Evaluating the Usability of Code Generation Tools Powered by Large Language Models},
year = {2022},
isbn = {9781450391566},
publisher = {Association for Computing Machinery},
address = {New York, NY, USA},
url = {https://doi.org/10.1145/3491101.3519665},
doi = {10.1145/3491101.3519665},
booktitle = {Extended Abstracts of the 2022 CHI Conference on Human Factors in Computing Systems},
articleno = {332},
numpages = {7},
location = {New Orleans, LA, USA},
series = {CHI EA '22}
}

@misc{like,
      title={What is it like to program with artificial intelligence?}, 
      author={Advait Sarkar and Andrew D. Gordon and Carina Negreanu and Christian Poelitz and Sruti Srinivasa Ragavan and Ben Zorn},
      year={2022},
      eprint={2208.06213},
      archivePrefix={arXiv},
      primaryClass={cs.HC},
      url={https://arxiv.org/abs/2208.06213}, 
}

@INPROCEEDINGS{validating,
  author={Tang, Ningzhi and Chen, Meng and Ning, Zheng and Bansal, Aakash and Huang, Yu and McMillan, Collin and Li, Toby Jia-Jun},
  booktitle={2024 IEEE Symposium on Visual Languages and Human-Centric Computing (VL/HCC)}, 
  title={Developer Behaviors in Validating and Repairing {LLM}-Generated Code Using {IDE} and Eye Tracking}, 
  year={2024},
  volume={},
  number={},
  pages={40-46},
  doi={10.1109/VL/HCC60511.2024.00015}}

@misc{autopilot,
      title={When {Copilot} Becomes Autopilot: Generative {AI}'s Critical Risk to Knowledge Work and a Critical Solution}, 
      author={Advait Sarkar and Xiaotong and Xu and Neil Toronto and Ian Drosos and Christian Poelitz},
      year={2024},
      eprint={2412.15030},
      archivePrefix={arXiv},
      primaryClass={cs.HC},
      url={https://arxiv.org/abs/2412.15030}, 
}

@inproceedings{critical,
author = {Lee, Hao-Ping (Hank) and Sarkar, Advait and Tankelevitch, Lev and Drosos, Ian and Rintel, Sean and Banks, Richard and Wilson, Nicholas},
title = {The Impact of Generative {AI} on Critical Thinking: Self-Reported Reductions in Cognitive Effort and Confidence Effects From a  Survey of Knowledge Workers},
booktitle = {Proceedings of the ACM CHI Conference on Human Factors in Computing Systems},
year = {2025},
month = {April},
publisher = {ACM},
url = {https://doi.org/10.1145/3706598.3713778},
doi = {10.1145/3706598.3713778}
}

@inproceedings{large-scale,
author = {Liang, Jenny T. and Yang, Chenyang and Myers, Brad A.},
title = {A Large-Scale Survey on the Usability of {AI} Programming Assistants: Successes and Challenges},
year = {2024},
isbn = {9798400702174},
publisher = {Association for Computing Machinery},
address = {New York, NY, USA},
url = {https://doi.org/10.1145/3597503.3608128},
doi = {10.1145/3597503.3608128},
booktitle = {Proceedings of the IEEE/ACM 46th International Conference on Software Engineering},
articleno = {52},
numpages = {13},
location = {Lisbon, Portugal},
series = {ICSE '24}
}

@article{ironies,
title = {Ironies of automation},
journal = {Automatica},
volume = {19},
number = {6},
pages = {775-779},
year = {1983},
issn = {0005-1098},
doi = {https://doi.org/10.1016/0005-1098(83)90046-8},
author = {Lisanne Bainbridge},
}

@misc{gaironies,
      title={Ironies of Generative {AI}: Understanding and mitigating productivity loss in human-{AI} interactions}, 
      author={Auste Simkute and Lev Tankelevitch and Viktor Kewenig and Ava Elizabeth Scott and Abigail Sellen and Sean Rintel},
      year={2024},
      eprint={2402.11364},
      archivePrefix={arXiv},
      howpublished = {arXiv},
      month = {February},
      primaryClass={cs.HC},
      url={https://arxiv.org/abs/2402.11364}, 
}

@article{bird,
author = {Bird, Christian and Ford, Denae and Zimmermann, Thomas and Forsgren, Nicole and Kalliamvakou, Eirini and Lowdermilk, Travis and Gazit, Idan},
title = {Taking Flight with {Copilot}: Early insights and opportunities of {AI}-powered pair-programming tools},
year = {2023},
issue_date = {November/December},
publisher = {Association for Computing Machinery},
address = {New York, NY, USA},
volume = {20},
number = {6},
issn = {1542-7730},
url = {https://doi.org/10.1145/3582083},
doi = {10.1145/3582083},
journal = {Queue},
month = jan,
pages = {35–57},
numpages = {23}
}

@inproceedings{between,
author = {Mozannar, Hussein and Bansal, Gagan and Fourney, Adam and Horvitz, Eric},
title = {Reading Between the Lines: Modeling User Behavior and Costs in {AI}-Assisted Programming},
year = {2024},
isbn = {9798400703300},
publisher = {Association for Computing Machinery},
address = {New York, NY, USA},
url = {https://doi.org/10.1145/3613904.3641936},
doi = {10.1145/3613904.3641936},
booktitle = {Proceedings of the 2024 CHI Conference on Human Factors in Computing Systems},
articleno = {142},
numpages = {16},
location = {Honolulu, HI, USA},
series = {CHI '24}
}

@article{comprehension,
author = {Maalej, Walid and Tiarks, Rebecca and Roehm, Tobias and Koschke, Rainer},
title = {On the Comprehension of Program Comprehension},
year = {2014},
issue_date = {August 2014},
publisher = {Association for Computing Machinery},
address = {New York, NY, USA},
volume = {23},
number = {4},
issn = {1049-331X},
url = {https://doi.org/10.1145/2622669},
doi = {10.1145/2622669},
journal = {ACM Trans. Softw. Eng. Methodol.},
month = sep,
articleno = {31},
numpages = {37},
}

@inproceedings{live,
author = {Ferdowsi, Kasra and Huang, Ruanqianqian (Lisa) and James, Michael B. and Polikarpova, Nadia and Lerner, Sorin},
title = {Validating {AI}-Generated Code with Live Programming},
year = {2024},
isbn = {9798400703300},
publisher = {Association for Computing Machinery},
address = {New York, NY, USA},
url = {https://doi.org/10.1145/3613904.3642495},
doi = {10.1145/3613904.3642495},
booktitle = {Proceedings of the 2024 CHI Conference on Human Factors in Computing Systems},
articleno = {143},
numpages = {8},
location = {Honolulu, HI, USA},
series = {CHI '24}
}

@inproceedings{understanding,
author = {Nam, Daye and Macvean, Andrew and Hellendoorn, Vincent and Vasilescu, Bogdan and Myers, Brad},
title = {Using an {LLM} to Help With Code Understanding},
year = {2024},
isbn = {9798400702174},
publisher = {Association for Computing Machinery},
address = {New York, NY, USA},
url = {https://doi.org/10.1145/3597503.3639187},
doi = {10.1145/3597503.3639187},
booktitle = {Proceedings of the IEEE/ACM 46th International Conference on Software Engineering},
articleno = {97},
numpages = {13},
location = {Lisbon, Portugal},
series = {ICSE '24}
}

@inproceedings{readable,
author = {Al Madi, Naser},
title = {How Readable is Model-generated Code? {E}xamining Readability and Visual Inspection of {GitHub} {Copilot}},
year = {2023},
isbn = {9781450394758},
publisher = {Association for Computing Machinery},
address = {New York, NY, USA},
url = {https://doi.org/10.1145/3551349.3560438},
doi = {10.1145/3551349.3560438},
booktitle = {Proceedings of the 37th IEEE/ACM International Conference on Automated Software Engineering},
articleno = {205},
numpages = {5},
location = {Rochester, MI, USA},
series = {ASE '22}
}

@article{its-weird,
author = {Prather, James and Reeves, Brent N. and Denny, Paul and Becker, Brett A. and Leinonen, Juho and Luxton-Reilly, Andrew and Powell, Garrett and Finnie-Ansley, James and Santos, Eddie Antonio},
title = {“{I}t’s Weird That it Knows What {I} Want”: Usability and Interactions with {Copilot} for Novice Programmers},
year = {2023},
issue_date = {February 2024},
publisher = {Association for Computing Machinery},
address = {New York, NY, USA},
volume = {31},
number = {1},
issn = {1073-0516},
url = {https://doi.org/10.1145/3617367},
doi = {10.1145/3617367},
journal = {ACM Trans. Comput.-Hum. Interact.},
month = nov,
articleno = {4},
numpages = {31},
}

@inproceedings{assistant,
author = {Ross, Steven I. and Martinez, Fernando and Houde, Stephanie and Muller, Michael and Weisz, Justin D.},
title = {The Programmer’s Assistant: Conversational Interaction with a Large Language Model for Software Development},
year = {2023},
isbn = {9798400701061},
publisher = {Association for Computing Machinery},
address = {New York, NY, USA},
url = {https://doi.org/10.1145/3581641.3584037},
doi = {10.1145/3581641.3584037},
booktitle = {Proceedings of the 28th International Conference on Intelligent User Interfaces},
pages = {491–514},
numpages = {24},
location = {Sydney, NSW, Australia},
series = {IUI '23}
}

@misc{li2024,
      title={{AI} Tool Use and Adoption in Software Development by Individuals and Organizations: A Grounded Theory Study}, 
      author={Ze Shi Li and Nowshin Nawar Arony and Ahmed Musa Awon and Daniela Damian and Bowen Xu},
      year={2024},
      eprint={2406.17325},
      archivePrefix={arXiv},
      primaryClass={cs.SE},
      url={https://arxiv.org/abs/2406.17325}, 
}

@misc{tightrope,
      title={Walking the Tightrope of {LLMs} for Software Development: A Practitioners' Perspective}, 
      author={Samuel Ferino and Rashina Hoda and John Grundy and Christoph Treude},
      year={2025},
      eprint={2511.06428},
      archivePrefix={arXiv},
      primaryClass={cs.SE},
      url={https://arxiv.org/abs/2511.06428}, 
}

@misc{skill-formation,
  author = {Shen, Judy Hanwen and Tamkin, Alex},
  title = {How {AI} Impacts Skill Formation},
  year = {2026},
  eprint = {2601.20245},
  archivePrefix = {arXiv},
  primaryClass = {cs.LG},
  eprinttype = {arxiv}
}

@book{glaser1,
  author    = {Glaser, Barney G. and Strauss, Anselm L.},
  title     = {The Discovery of Grounded Theory: Strategies for Qualitative Research},
  year      = {1967},
  publisher = {Aldine de Gruyter},
  address   = {New York}
}

@book{
glaser2,
title = {Theoretical sensitivity : advances in the methodology of grounded theory},
author = {Glaser, Barney G.},
address = {Mill Valley (Calif.)},
publisher = {Sociology Press},
year = {1978},
}

@book{strauss,
title = {Basics of qualitative research : techniques and procedures for developing grounded theory},
author = {Corbin, Juliet and Strauss, Anselm L.},
address = {Los Angeles, Calif.},
publisher = {Sage},
year = {2015},
edition = {4th},
}

\appendix

\end{document}